\documentclass[3p,times,procedia]{elsarticle}
\usepackage{nupha_ecrc}
\usepackage[]{amsmath}
\def\bra{\langle}
\def\ket{\rangle}

\volume{00}

\firstpage{1}

\journalname{Nuclear Physics A}

\runauth{}

\jid{nupha}

\jnltitlelogo{Nuclear Physics A}

\usepackage{amssymb}

\usepackage[figuresright]{rotating}
\usepackage{url}
\usepackage{hyperref}

\begin{document}

\begin{frontmatter}



\dochead{XXVIIth International Conference on Ultrarelativistic Nucleus-Nucleus Collisions\\ (Quark Matter 2018)}

\title{Nonlinear coupling of flow harmonics in heavy-ion collisions}


\author[label1]{Giuliano Giacalone}



\address[label1]{Institut de physique th\'eorique, Universit\'e Paris Saclay, CNRS, CEA, F-91191 Gif-sur-Yvette, France}


\begin{abstract}
Anisotropic flow coefficients beyond triangular flow receive important contributions from lower-order harmonics through nonlinear coupling.
We present a theoretical framework which allows one to quantify the contribution induced by such nonlinear couplings to any flow harmonic of any order.
We show the effectiveness of this formalism through an application to hexagonal flow, $V_6$. 
We study, in particular, the coupling of $V_6$ to triangular flow, $V_3$, in Pb+Pb collisions at $~\sqrt[]{s}=2.76$ TeV, using both Large Hadron Collider data and event-by-event hydrodynamic calculations.
\end{abstract}


\end{frontmatter}


\section{Introduction}
\label{sec:1}
Particles detected in the final states of relativistic nucleus-nucleus collisions present a spectacular degree of azimuthal anisotropy, in the sense that if we perform a harmonic analysis of their azimuthal directions via Fourier decomposition:
\begin{equation}
\label{eq:vn}
\frac{dN}{p_{\rm t} dp_{\rm t}d\phi} = \frac{dN}{p_{\rm t} dp_{\rm t}} \frac{1}{2\pi} \sum_{n=-\infty}^{\infty} V_n e^{in\phi} ,\hspace{40pt}V_{-n}=V_n^*,
\end{equation}
the coefficients $V_n$ turn out to be nonzero.
This observation has a natural explanation in a hydrodynamic framework~\cite{Yan:2017ivm}, according to which particles are emitted to the final state after the hydrodynamic expansion of a fluidlike medium created in the collision zone.
In this framework, Fourier anisotropy is of geometric origin~\cite{Teaney:2010vd}, as the two most prominent coefficients of the spectrum, $V_2$ and $V_3$, emerge as a response of the fluid to the elliptic and triangular anisotropy, respectively, of its initial geometry~\cite{Niemi:2015qia}.

The origin of harmonics with $n>3$ is instead more complicated, as shown by the following argument~\cite{Teaney:2012ke}.
Anisotropic flow in the medium yields a transverse field of velocity amenable to a an expansion in harmonics:
\begin{equation}
\label{eq:ut}
u_t(\phi_u)=u_{t,0}+u_{t,2} \cos(2\phi_u)+u_{t,4} \cos(4\phi_u)+ \ldots + \rm {odd~terms.}~
\end{equation}
At freeze-out, since the momentum density of emitted particles is close to an equilibrium distribution, dubbing $\phi_p$ the azimuthal orientation of transverse momentum, we can use Eq.~(\ref{eq:ut}) to write:
\begin{equation}
\label{eq:exp}
\frac{dN}{p_{\rm t} dp_{\rm t}d\phi_p} \propto \exp \biggl( -\frac{\vec u_t\cdot \vec p_t}{T} \biggr) \simeq \exp \biggl( -\frac{p_t}{T}u_{t,0}\cos (\phi_u - \phi_p) \biggr) \biggl[ 1+\frac{p_t}{T}u_{t,2}\cos(2 \phi_u)  +\frac{1}{2} \biggl( \frac{p_t}{T}u_{t,2}\cos(2 \phi_u) \biggr)^2 +\ldots\biggr].
\end{equation}
Now, using the saddle-point approximation explained in Ref.~\cite{Borghini:2005kd}, we set $\phi_u=\phi_p$ inside the square brackets on the right hand side of Eq.~(\ref{eq:exp}), and we obtain that the angular anisotropy of the velocity field contributes to all orders in the harmonic spectrum of the final-state momenta.
For instance, the term involving $\bigl(\cos(2\phi_u)\bigr)^2$ on the rhs generates a $V_2^2$ in the spectrum of the lhs, which eventually contributes to the overall fourth harmonic, $V_4$.
This feature is crucial in the context of heavy-ion collisions. 
Since the measured spectrum is strongly ordered, i.e., $V_2>V_3>V_4>\ldots$~, one naturally expects harmonics of order $n>3$ to originate almost entirely from their coupling to $V_2$ and $V_3$.
In the following, we present a theoretical formalism that allows one to compute the contributions to a given flow coefficient, $V_n$, induced by harmonics of lower order.

\section{A framework of nonlinear coupling of flow harmonics}
\label{sec:2}
The formalism was introduced in Ref.~\cite{Giacalone:2018wpp}.
For a given Fourier harmonic, $V$, in a given class of impact parameter (or centrality), we perform a decomposition of the kind:
\begin{equation}
\label{decomp}
V=\sum_{k=1}^p\chi_k W_k + U,
\end{equation}
where $W_k$ are combinations of lower-order harmonics that contribute to $V$, and $U$ is a complex quantity which we define by requiring
\begin{equation}
\label{uncorrelated2}
\langle W_k^* U\rangle=0, \hspace{5pt}\forall k,
\end{equation}
where $\bra\ldots\ket$ is an average over events. 
The quantities we are eventually interested in obtaining are the coefficients, $\chi_k$, the so-called nonlinear response coefficients, that allow us to quantify the coupling between $V$ and the lower-order harmonics, $W$.
Using Eq.~(\ref{uncorrelated2}), we can write down the following identity
\begin{equation}
\label{system}
\langle W_j^*V\rangle=\sum_{k=1}^p \chi_k \langle W_j^*W_k\rangle,
\end{equation}
which corresponds to a linear system of $p$ equations for $p$ constants.
Now, defining the following $p\times p$ hermitian matrix
\begin{equation}
\label{defsigma}
\Sigma_{jk}\equiv  \langle W_j^*W_k\rangle, 
\end{equation}
the previous system, Eq~(\ref{system}), can be rewritten in matrix form:
\begin{equation}
\label{systemm}
M=\Sigma X,
\end{equation}
where $M$ is a $p$-vector with components $\langle W_j^*V\rangle$, and $X$ a vector containing the $p$ coupling constants.
Therefore, if $\Sigma$ and $M$ are known, the coupling coefficients come from straightforward matrix inversion:
\begin{equation}
\label{eq:chi}
X=\Sigma^{-1} M.
\end{equation}
Equation~(\ref{eq:chi}) is our main result.
In the remainder of this paper, we perform an application of this formalism to the extraction of the coupling coefficients characterizing hexagonal flow, $V=V_6$.

\section{Application to hexagonal flow}
\label{sec:3}
The nonlinear mode expansion of hexagonal flow reads\footnote{We always neglect contributions involving the first harmonic, $V_1$, which is subleading.}~\cite{Qian:2016fpi}.
\begin{equation}
\label{v6decomposition}
V_6=\chi_{62}V_2^3+\chi_{63}V_3^2+\chi_{624}V_2U_4+U_6,
\end{equation}
where $U_4=V_4-\chi_{42}V_2^2$, with $\bra V_2^2 U_4^* \ket=0$. 
The matrix in Eq.~(\ref{defsigma}) is therefore given by
\begin{equation}
\label{matrixv6}
\Sigma^{(6)} \equiv \Sigma = 
\begin{pmatrix}
\bra v_2^6 \ket  & \bra (V_2^*)^3 V_3^2 \ket & \bra v_2^2 U_4 (V_2^2)^* \ket \cr
\bra V_2^3 (V_3^*)^2 \ket  & \bra v_3^4 \ket &\bra (V_3^2)^* U_4 V_2 \ket  \cr
\bra v_2^2 U_4^* V_2^2 \ket  &  \bra V_3^2 U_4^* V_2^* \ket & \bra u_4^2v_2^2 \ket 
\end{pmatrix},
\end{equation}
whereas the vector $M$ reads
\begin{equation}
\label{vectorv6}
M=\biggl(\bra (V_2^*)^3 V_6\ket,~\bra (V_3^*)^2 V_6\ket,~\bra V_2^*U_4^* V_6\ket\biggr),
\end{equation}
with the notation $v_n=|V_n|$ and $u_n=|U_n|$.
We neglect effects of parity violation, and we take $\Sigma^{(6)}$ to be real.

\begin{figure}[t!]
\centering
\includegraphics[width=.95\linewidth]{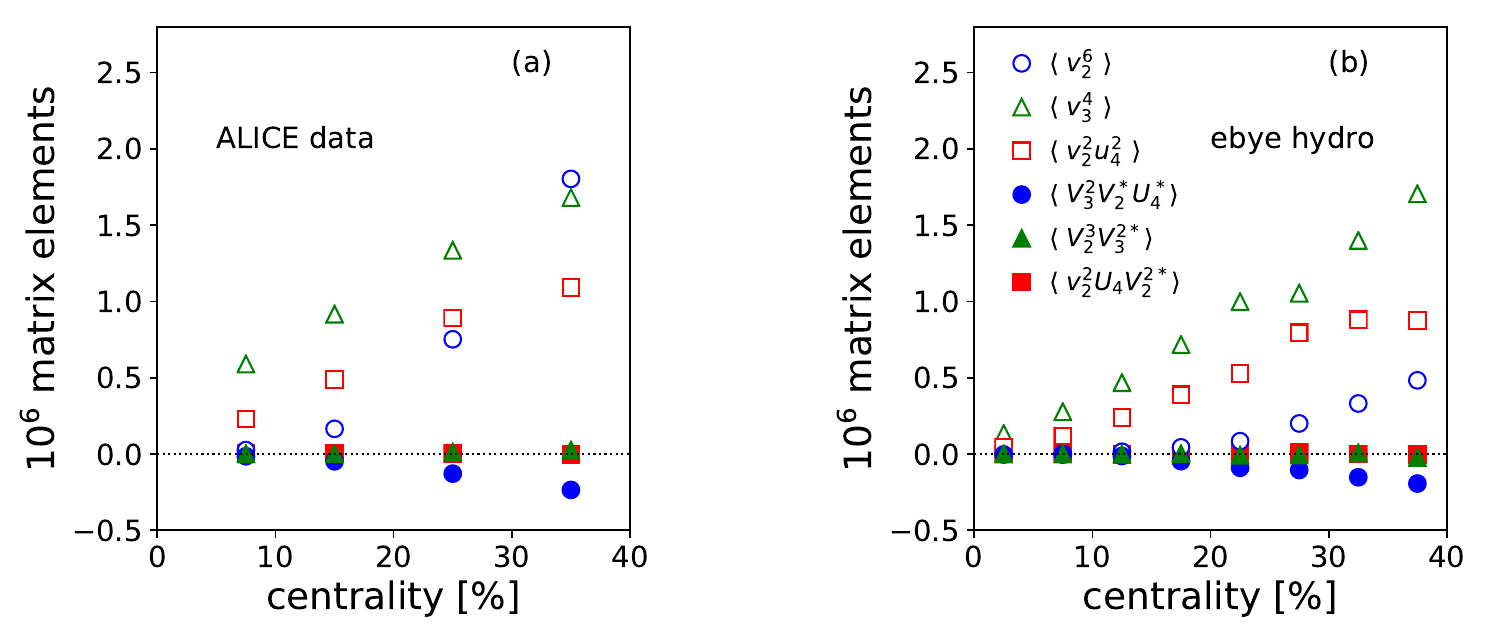}
\caption{Elements of $\Sigma^{(6)}$ from ALICE data [panel (a)], and from hydrodynamic calculations [panel (b)].}
\label{fig:1}
\end{figure}

We extract both $\Sigma^{(6)}$ and $M$ from experimental data on Pb+Pb collisions at $~\sqrt[]{s}=2.76$~GeV.
Almost all the required moments can be obtained from ALICE data~\cite{ALICE:2016kpq,Acharya:2017zfg}, though one has to resort to ATLAS data~\cite{Aad:2014fla} on event-plane correlations for two- and three-plane correlators.
We refer to Ref.~\cite{Giacalone:2018wpp} for a thorough description of the extraction procedure.
It is instructive to look at the centrality dependence of the elements of $\Sigma^{(6)}$, displayed in Fig.~\ref{fig:1}(a).
The matrix is to a good approximation diagonal, with only a tiny off-diagonal term provided by the three-plane correlator $\bra V_3^2 U_4^* V_2^* \ket$.
Application of Eq.~(\ref{eq:chi}) yields, eventually, the coupling coefficients of $V_6$ as function of centrality percentile.
The resulting coefficients of $V_6$ are reported in Ref.~\cite{Giacalone:2018wpp}.

\begin{figure}[t!]
\centering
\includegraphics[width=.6\linewidth]{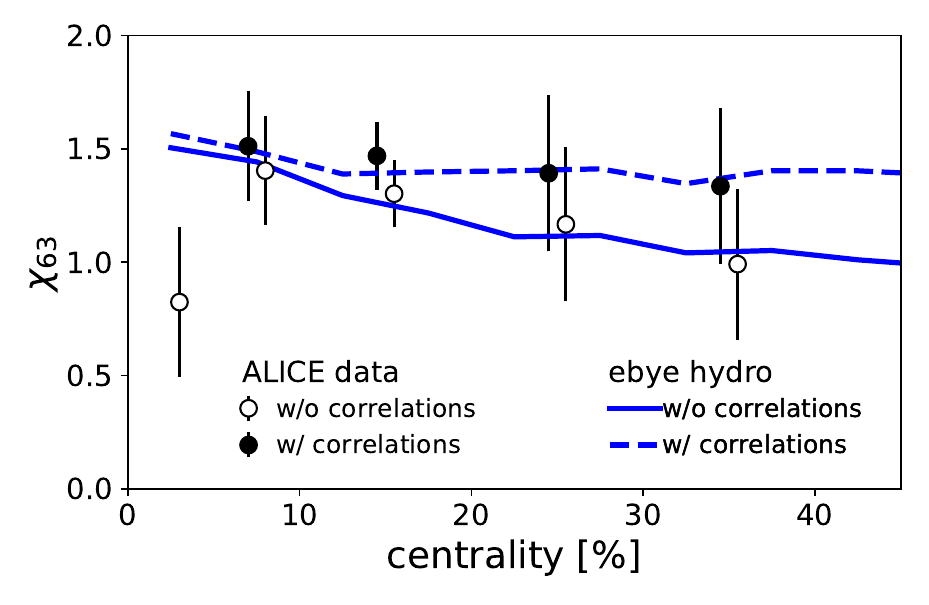}
\caption{Coefficient $\chi_{63}$ from experimental data (symbols) and hydrodynamic calculations (lines). Different symbol and line styles indicate the inclusion/exclusion of the off-diagonal elements of $\Sigma^{(6)}$ in the calculation.}
\label{fig:2}
\end{figure}

For reasons that will appear clear in a moment, here we focus solely on the coupling between $V_6$ and $V_3$, $\chi_{63}$, which is displayed as full symbols in Fig.~\ref{fig:2}.
Results are excellent, in the sense that the coefficient has a mild dependence on the centrality percentile, as one expects from generic arguments~\cite{Yan:2015jma}.
Now, as pointed out in Ref.~\cite{Giacalone:2018wpp}, $\chi_{63}$ provides a clear example of the effectiveness of the matrix formalism outlined above.
Previous analyses of nonlinear flow modes~\cite{Qian:2016fpi,Acharya:2017zfg,Yan:2015jma,Qian:2017ier,Zhao:2017yhj}  have been carried out neglecting the mutual correlations between nonlinear terms in Eq.~(\ref{v6decomposition}), or in our language, neglecting all off-diagonal terms of $\Sigma^{(6)}$: 
\begin{equation}
\bra V_2^3 (V_3^2)^* \ket = 0, \hspace{40pt} \bra v_2^2 U_4^* V_2^2  \ket = 0, \hspace{40pt} \bra V_3^2 U_4^* V_2^*  \ket =0.
\end{equation}
With this approximation, experimental data on $\chi_{63}$ are given by the white symbols in Fig.~\ref{fig:2}, and turn out to present a much stronger centrality dependence, which indicates a worse determination of the true value of the coupling coefficient.
Therefore, the nonzero off-diagonal element of $\Sigma^{(6)}$, though small, leaves a visible signature in $\chi_{63}$.
Such effect is fully taken into account in the matrix formalism, where no approximation is made concerning the correlations that contribute to $V_6$.

We go a little beyond the purely data-driven analysis of Ref.~\citep{Giacalone:2018wpp}, and check whether the influence of a nonzero $\bra V_3^2 U_4^* V_2^* \ket$ on the coefficient $\chi_{63}$ is an actual prediction of hydrodynamics.
To this aim, we run event-by-event (ebye) viscous hydrodynamic simulations of Pb+Pb collisions at $~\sqrt[]{s}=2.76$~TeV, using the code v-USPHYDRO\cite{Noronha-Hostler:2013gga,Noronha-Hostler:2014dqa}, with the same parameter setup as in Ref.~\cite{Giacalone:2016afq}.
We first compute $\Sigma^{(6)}$, whose elements are shown in Fig.~\ref{fig:1}(b).
Hydrodynamics captures the generic feature displayed by the experimental data: The matrix is essentially diagonal, with one tiny off-diagonal element given by $\bra V_3^2 U_4^* V_2^* \ket$.
Computing $M$, and using Eq.~(\ref{eq:chi}), we eventually extract the coefficient $\chi_{63}$.
Results are shown in Fig.~\ref{fig:2}, and are in good agreement with the data.
We compute the coefficient both including the full matrix $\Sigma^{(6)}$ (dashed line), and neglecting off-diagonal terms (solid line).
We note that the centrality dependence of $\chi_{63}$ becomes much weaker if the full matrix is included, as observed for the experimental data.
We conclude that hydrodynamics correctly captures the effect of a nonzero off-diagonal element of $\Sigma^{(6)}$ on the determination of $\chi_{63}$.

\section{Conclusive remarks and outlook}
We expect high-statistics LHC2 data, combined with our theoretical formalism, to lead to a rich phenomenology of nonlinear harmonic coupling in the near future.
The coefficients $\chi$ appear to depend mildly on centrality, initial conditions, and in general on medium properties during the hydrodynamic phase. 
They probe, on the other hand, \textit{some} generic feature of the late-time dynamics of the quark-gluon plasma. 
What we are missing is precisely a solid understanding of what such physics is.
Clarifying this issue in view of upcoming measurements is an exciting prospect for future theoretical studies.

We thank Jacquelyn Noronha-Hostler for providing us with the results of v-USPHYDRO calculations.

\bibliographystyle{elsarticle-num}
\bibliography{biblio}

\begin{thebibliography}{10}
\expandafter\ifx\csname url\endcsname\relax
  \def\url#1{\texttt{#1}}\fi
\expandafter\ifx\csname urlprefix\endcsname\relax\def\urlprefix{URL }\fi
\expandafter\ifx\csname href\endcsname\relax
  \def\href#1#2{#2} \def\path#1{#1}\fi

\bibitem{Yan:2017ivm}
L.~Yan, Chin. Phys. C42~(4) (2018) 042001.
\newblock \href {http://arxiv.org/abs/1712.04580} {\path{arXiv:1712.04580}}.

\bibitem{Teaney:2010vd}
D.~Teaney, L.~Yan, Phys. Rev. C83 (2011) 064904.
\newblock \href {http://arxiv.org/abs/1010.1876} {\path{arXiv:1010.1876}}.

\bibitem{Niemi:2015qia}
H.~Niemi, K.~J. Eskola, R.~Paatelainen, Phys. Rev. C93~(2) (2016) 024907.
\newblock \href {http://arxiv.org/abs/1505.02677} {\path{arXiv:1505.02677}}.

\bibitem{Teaney:2012ke}
D.~Teaney, L.~Yan, Phys. Rev. C86 (2012) 044908.
\newblock \href {http://arxiv.org/abs/1206.1905} {\path{arXiv:1206.1905}}.

\bibitem{Borghini:2005kd}
N.~Borghini, J.-Y. Ollitrault, Phys. Lett. B642 (2006) 227--231.
\newblock \href {http://arxiv.org/abs/nucl-th/0506045}
  {\path{arXiv:nucl-th/0506045}}.

\bibitem{Giacalone:2018wpp}
G.~Giacalone, L.~Yan, J.-Y. Ollitrault, Phys. Rev. C97~(5) (2018) 054905.
\newblock \href {http://arxiv.org/abs/1803.00253} {\path{arXiv:1803.00253}}.

\bibitem{Qian:2016fpi}
J.~Qian, U.~W. Heinz, J.~Liu, Phys. Rev. C93~(6) (2016) 064901.
\newblock \href {http://arxiv.org/abs/1602.02813} {\path{arXiv:1602.02813}}.

\bibitem{ALICE:2016kpq}
J.~Adam, et~al., Phys. Rev. Lett. 117 (2016) 182301.
\newblock \href {http://arxiv.org/abs/1604.07663} {\path{arXiv:1604.07663}}.

\bibitem{Acharya:2017zfg}
S.~Acharya, et~al., Phys. Lett. B773 (2017) 68--80.
\newblock \href {http://arxiv.org/abs/1705.04377} {\path{arXiv:1705.04377}}.

\bibitem{Aad:2014fla}
G.~Aad, et~al., Phys. Rev. C90~(2) (2014) 024905.
\newblock \href {http://arxiv.org/abs/1403.0489} {\path{arXiv:1403.0489}}.

\bibitem{Yan:2015jma}
L.~Yan, J.-Y. Ollitrault, Phys. Lett. B744 (2015) 82--87.
\newblock \href {http://arxiv.org/abs/1502.02502} {\path{arXiv:1502.02502}}.

\bibitem{Qian:2017ier}
J.~Qian, U.~Heinz, R.~He, L.~Huo, Phys. Rev. C95~(5) (2017) 054908.
\newblock \href {http://arxiv.org/abs/1703.04077} {\path{arXiv:1703.04077}}.

\bibitem{Zhao:2017yhj}
W.~Zhao, H.-j. Xu, H.~Song, Eur. Phys. J. C77~(9) (2017) 645.
\newblock \href {http://arxiv.org/abs/1703.10792} {\path{arXiv:1703.10792}}.

\bibitem{Noronha-Hostler:2013gga}
J.~Noronha-Hostler, G.~S. Denicol, J.~Noronha, R.~P.~G. Andrade, F.~Grassi,
  Phys. Rev. C88~(4) (2013) 044916.
\newblock \href {http://arxiv.org/abs/1305.1981} {\path{arXiv:1305.1981}}.

\bibitem{Noronha-Hostler:2014dqa}
J.~Noronha-Hostler, J.~Noronha, F.~Grassi, Phys. Rev. C90~(3) (2014) 034907.
\newblock \href {http://arxiv.org/abs/1406.3333} {\path{arXiv:1406.3333}}.

\bibitem{Giacalone:2016afq}
G.~Giacalone, L.~Yan, J.~Noronha-Hostler, J.-Y. Ollitrault, Phys. Rev. C94~(1)
  (2016) 014906.
\newblock \href {http://arxiv.org/abs/1605.08303} {\path{arXiv:1605.08303}}.

\end{thebibliography}







\end{document}